\documentclass[twocolumn,english,reprint,prl]{revtex4-1}

\usepackage[T1]{fontenc}
\usepackage[latin9]{inputenc}
\usepackage{color}
\usepackage{bm}
\usepackage{amstext}
\usepackage{amssymb}
\usepackage{graphicx}
\usepackage{booktabs}
\usepackage{tabularx}
\usepackage{amsmath}
\makeatletter

\usepackage{babel}

\makeatother

\begin{document}
\author{Giacomo Torlai}
\affiliation{Department of Physics and Astronomy, University of Waterloo, Ontario N2L 3G1, Canada,}
\affiliation{Perimeter Institute for Theoretical Physics, Waterloo, Ontario N2L 2Y5, Canada,}
\author{Roger G. Melko}
\affiliation{Department of Physics and Astronomy, University of Waterloo, Ontario N2L 3G1, Canada,}
\affiliation{Perimeter Institute for Theoretical Physics, Waterloo, Ontario N2L 2Y5, Canada,}
\title{Latent Space Purification via Neural Density Operators}
\begin{abstract}
Machine learning is actively being explored for its potential to design, validate, and even hybridize with near-term quantum devices. A central question is whether neural networks can provide a tractable representation of a given quantum state of interest.  When true, stochastic neural networks can be employed for many unsupervised tasks, including generative modeling and state tomography. However, to be applicable for real experiments 
such methods must be able to encode quantum mixed states. Here, we parametrize a density matrix based on a 
restricted Boltzmann machine that is capable of purifying a mixed state through auxiliary degrees of freedom embedded in the latent space of
its hidden units. We implement the algorithm numerically and use it to perform tomography on some typical states of entangled photons, achieving fidelities competitive with standard 
techniques.
\end{abstract}
\maketitle
\selectlanguage{english}
{\it Introduction.}  
Quantum materials, matter, and devices have highly complex features that can make describing the correlations between particles challenging, even for the world's most powerful computers.  Classical algorithms have been instrumental in the design and characterization of quantum systems, ranging from the scale of few-body molecules and devices, up to many-body atomic and condensed matter. However, to be successful in reproducing the behavior of even a small number of qubits, such algorithms may require a very large amount of classical resources,
 a fact which presents a continuing challenge for modern quantum sciences.

For the moderately small number of qubits presently manipulated in laboratories, one can imagine optimizing a classical model with the most efficient 
means currently available to today's conventional hardware, in such a way that a faithful representation of a generic quantum state is produced.
Machine learning (ML) of graphical models, based on neural networks with a latent space formed by hidden variables, provides one of the most practical routes to achieving this. Here, the task of reducing the dimensionality of the Hilbert space is conceptually similar to identifying relevant (low-dimensional) features hidden within a higher-dimensional data set~\cite{hinton_reducing_2006}.
Modern algorithms for ML are sufficiently advanced to allow neural network models to be 
learned in a reasonable time, from real data sets obtained from measurements of present-day 
experimental or synthetic quantum systems.

Recently, a number of authors have demonstrated that
a type of stochastic neural network, called a restricted Boltzmann machine (RBM), can be used to capture various properties of many-body systems. These include the thermodynamics of spin models~\cite{torlai_learning_2016}, ground-state and dynamical properties of interacting quantum spins~\cite{Carleo2017}, quantum nonlocality~\cite{Deng2017}, and quantum error correction~\cite{torlai_neural_2016} for example. The underlying representational power of such networks is currently 
under intense theoretical investigation~\cite{deng_exact_2016,deng_quantum_2017,Huang2017,Cheng_Chen_Wang,Bell_RBM,Glasser2018}. 
Numerically, RBMs have been successfully trained, using standard ML techniques, to faithfully represent a variety of quantum many-body wave functions, for numbers of qubits ranging into the hundreds.  In Ref.~\cite{torlai_Tomo}, it was demonstrated how an
RBM with hidden units could be used to perform quantum-state tomography, by learning to represent a pure 
many-body wave function within its network parameters, trained from a finite-size set 
of measurement data.

However, in realistic applications in the laboratory, quantum states are difficult to isolate, and are often entangled to the environment.
Hence, the purity of the underlying system cannot be assumed, and 
tomography must generally be performed on states with unknown mixing.
In this Letter we extend the concept of pure-state tomography with RBMs \cite{torlai_Tomo},
to the more general class of mixed states described by density matrices. The resulting graphical model, which we call a neural density operator (NDO), is obtained by purifying the mixed state of the physical system through additional auxiliary degrees of freedom, embedded in the latent space of hidden variables in the neural network. 
Upon tracing out the hidden variables, the network becomes a representation of the density matrix.
We derive a generalization of the most effective known training algorithm for RBMs, which minimizes a 
Kullbach-Leibler divergence through
contrastive divergence.
Then, we implement our algorithm numerically, and demonstrate that it is able to
reconstruct the density matrix of an unknown quantum state by training a NDO on a set of measurements.  
As an example, we illustrate the state reconstruction algorithm on real experimental data, for a simple case of two entangled photons.

{\it Neural density operators.}  We consider the state of a quantum system comprising $N$ degrees of freedom, characterized by a density operator $\bm{\rho}$ with matrix elements $\rho(\bm{\sigma},\bm{\sigma}^\prime)=\langle\bm{\sigma}|\bm{\rho}|\bm{\sigma}^\prime\rangle$, in an (arbitrary) reference basis $\bm{\sigma}\equiv(\sigma_1,\dots,\sigma_N)$. For simplicity, we restrict ourselves to the case of a two-dimensional local Hilbert space $\sigma_j=\{0,1\}$ (e.g. $\frac{1}{2}$-spins, hard-core bosons, qubits, etc.). When the system is in a pure state, the density operator assumes the simple form $\bm{\rho}=|\psi\rangle\langle\psi|$ given by the wave function $|\psi\rangle=\sum_{\bm{\sigma}}\psi(\bm{\sigma})|\bm{\sigma}\rangle$. In this case, as shown by Carleo and Troyer~\cite{Carleo2017}, any quantum state has a RBM representation $|\psi_{\bm{\theta}}\rangle$, where the wave function is encoded into a set of internal parameters $\bm{\theta}$ of a neural network (the number of which generally grows exponentially for a generic quantum state).
The encoded state is then a highly nonlinear function which returns a complex-valued coefficient $\psi_{\bm{\theta}}(\bm{\sigma})$, for any input state $|\bm{\sigma}\rangle$. The optimal set of parameters which best approximates the wave function is found by training the neural network with a ``learning'' procedure. 
For example, this could be the variational minimization of the total energy \cite{Carleo2017,Nomura2017}. Alternatively, training can occur via standard machine learning procedures, if an appropriate data set is available \cite{torlai_Tomo}.
Depending on the complexity of the state to be encoded, different numbers of network parameters will be required, which naturally 
quantifies a convergence parameter for the algorithm.

\begin{figure}[t]
\noindent \centering{}\includegraphics[width=0.75\columnwidth]{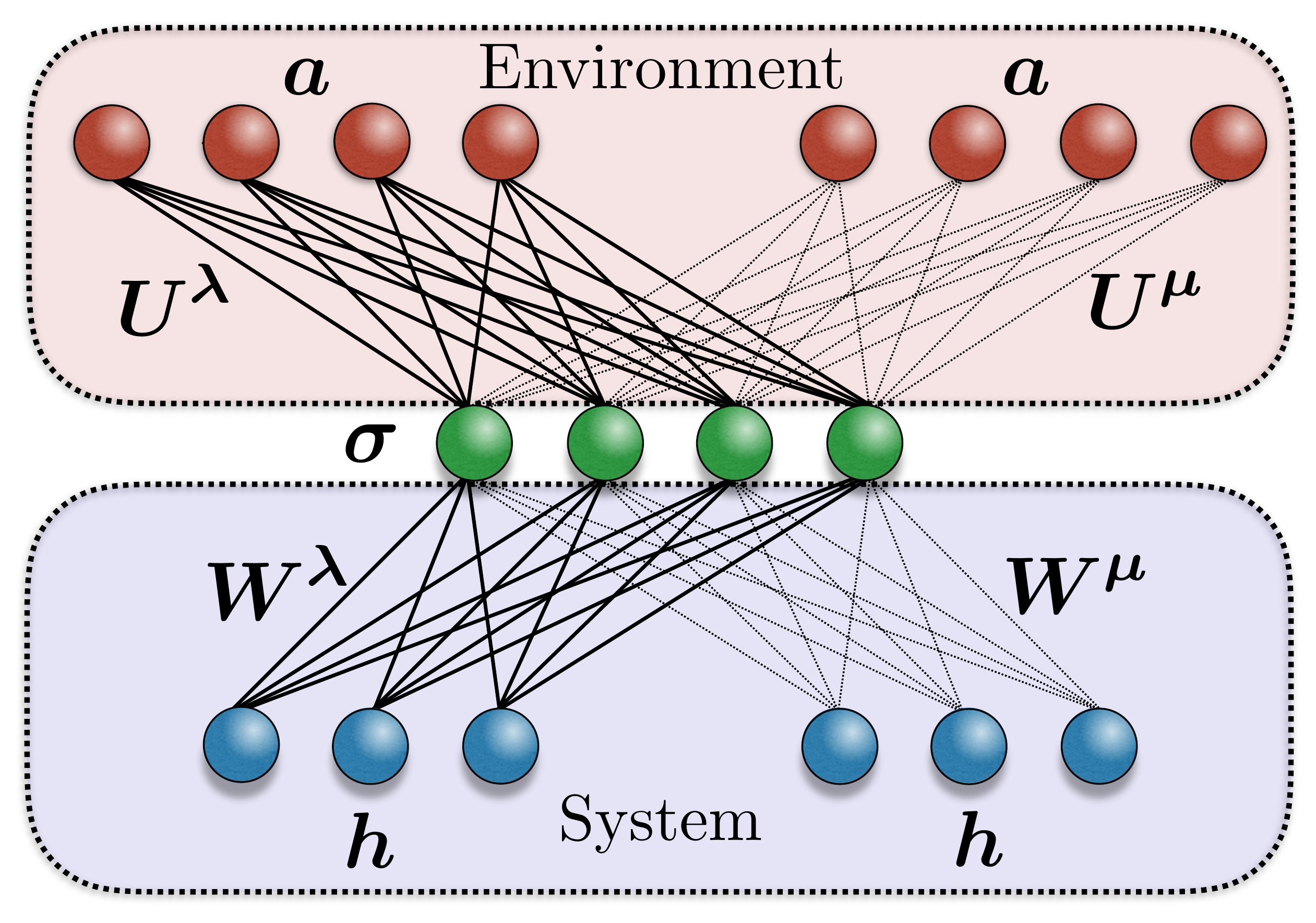}
\caption{Graphical representation of the neural density operator. The visible layer (green) encodes the state of the physical system $\bm{\sigma}$, while the other two layers are used to describe the mixing due to the environment (red), and to capture the correlations between the physical degrees of freedom (blue).}
\label{Fig1} 
\end{figure}

In analogy to this, we define the NDO as a mapping $\bm{\rho}_{\bm{\theta}}$ that, given two input states $|\bm{\sigma}\rangle$ and $|\bm{\sigma}^\prime\rangle$, returns the matrix element $\rho_{\bm{\theta}}(\bm{\sigma},\bm{\sigma}^\prime)$. For a NDO to describe a physical state, its matrix representation must have unit trace $\text{Tr}_{\bm{\sigma}}\{\bm{\rho_{\bm{\theta}}}\}=1$, must be Hermitian $\bm{\rho}_{\bm{\theta}}=\bm{\rho}_{\bm{\theta}}^\dagger$, and must be positive semidefinite $\langle \bm{x}|\bm{\rho}_{\bm{\theta}}|\bm{x}\rangle\ge0\:\:\forall|\bm{x}\rangle$. 
These constraints can be satisfied by constructing the NDO from the purification of its Hilbert space with a system of $n_a$ 
auxiliary degrees of freedom $\bm{a}=(a_1,\dots,a_{n_a})$, so that its composite state  $\bm{\rho^{\bm{\sigma}\oplus\bm{a}}_{\bm{\theta}}}$ is pure, and therefore $\bm{\rho^{\bm{\sigma}\oplus\bm{a}}_{\bm{\theta}}}=|\psi_{\bm{\theta}}\rangle\langle\psi_{\bm{\theta}}|$, with a neural network wave function  $|\psi_{\bm{\theta}}\rangle=\sum_{\bm{\sigma}\bm{a}}\psi_{\bm{\theta}}(\bm{\sigma},\bm{a})|\bm{\sigma}\rangle\otimes|\bm{a}\rangle$. 
The NDO is then simply obtained by tracing out the auxiliary system $\bm{\rho_{\bm{\theta}}}=\text{Tr}_{\bm{a}}\{|\psi_{\bm{\theta}}\rangle\langle\psi_{\bm{\theta}}|\}$, obtaining the density matrix
\begin{equation}
\rho_{\bm{\theta}}(\bm{\sigma},\bm{\sigma}^\prime)=\sum_{\bm{a}}\psi_{\bm{\theta}}(\bm{\sigma},\bm{a})\psi^*_{\bm{\theta}}(\bm{\sigma}^\prime,\bm{a}) .
\label{NDO}
\end{equation}
While the nature of the auxiliary system is arbitrary, a RBM provides a very convenient method for encoding both the physical and auxiliary degrees of freedom. A standard RBM contains two layers of stochastic binary units, a visible or physical layer, and a hidden or latent layer $\bm{h}$. The two layers are connected by a set of weighted edges, and each unit is also coupled to an external field (or bias). Here, we embed the auxiliary units used for the purification in the hidden layer of the neural network, which is thus enlarged to $(\bm{h},\bm{a})$. The RBM associates to this graph structure a Boltzmann probability distribution $p_{\bm{\theta}}(\bm{\sigma},\bm{a},\bm{h})$, where the network parameters are $\bm{\theta}=\{\bm{W_\theta},\bm{U_\theta},\bm{b_\theta},\bm{c_\theta},\bm{d_\theta}\}$ (see Fig.~\ref{Fig1}). Thee distribution describing the composite (pure) system is obtained by integrating out the hidden variables $\bm{h}$:
\begin{equation}
\begin{split}
p_{\bm{\theta}}(\bm{\sigma},\bm{a})=e^{\sum_i\log(1+e^{\bm{W}^{[i]}_{\bm{\theta}}\bm{\sigma}+\bm{c}_{\bm{\theta}}^{[i]}})+\bm{a}^\top\bm{U}_{\bm{\theta}}\bm{\sigma}+\bm{b}_{\bm{\theta}}^\top\bm{\sigma}+\bm{d}_{\bm{\theta}}^\top\bm{a}}
\end{split}
\end{equation}
with $\bm{W}^{[i]}_{\bm{\theta}}$ and $\bm{c}_{\bm{\theta}}^{[i]}$ begin the $i$th rows of the weight matrix and hidden field. We define the quantum state of the composite system using two sets of parameters $\bm{\theta}=(\bm{\lambda},\bm{\mu})$ describing amplitudes and phases respectively:
\begin{equation}
\psi_{\bm{\lambda\mu}}(\bm{\sigma},\bm{a})=Z_{\bm{\lambda}}^{-\frac{1}{2}}\sqrt{p_{\bm{\lambda}}(\bm{\sigma},\bm{a})}e^{i\phi_{\bm{\mu}}(\bm{\sigma},\bm{a})}
\end{equation}
where $\phi_{\bm{\mu}}(\bm{\sigma},\bm{a})=\log p_{\bm{\mu}}(\bm{\sigma},\bm{a})/2$ and $Z_{\bm{\lambda}}=\sum_{\bm{\sigma}\bm{a}}p_{\bm{\lambda}}(\bm{\sigma},\bm{a})$ is a constant enforcing normalization.

Since the auxiliary units are embedded in the latent space of the network, we can perform the summation in Eq.~\eqref{NDO} exactly, obtaining $\bm{\rho}_{\bm{\lambda\mu}}=Z_{\bm{\lambda}}^{-1}\bm{\tilde{\rho}}_{\bm{\lambda\mu}}$ with unnormalized matrix elements 
\begin{equation}
\tilde{\rho}_{\bm{\lambda\mu}}(\bm{\sigma},\bm{\sigma}^\prime)=e^{\Gamma_{\bm{\lambda}}^{[+]}(\bm{\sigma},\bm{\sigma}^\prime)+i\Gamma_{\bm{\mu}}^{[-]}(\bm{\sigma},\bm{\sigma}^\prime)+\Pi_{\bm{\lambda}\bm{\mu}}(\bm{\sigma},\bm{\sigma}^\prime)}
\end{equation}
Here we have introduced the matrices
\begin{equation}
\begin{split}
\Gamma_{\bm{\theta}}^{[\pm]}(\bm{\sigma},\bm{\sigma}^\prime)&=\frac{1}{2}\bigg[\sum_i\log(1+e^{\bm{W}^{[i]}_{\bm{\theta}}\bm{\sigma}+\bm{c}^{[i]}_{\bm{\theta}}})\\
&\pm\sum_i\log(1+e^{\bm{W}^{[i]}_{\bm{\theta}}\bm{\sigma}^\prime+\bm{c}^{[i]}_{\bm{\theta}}})+\bm{b}_{\bm{\theta}}^\top(\bm{\sigma}\pm\bm{\sigma}^\prime)\bigg]
\end{split}
\end{equation}
and
\begin{equation}
\begin{split}
\Pi_{\bm{\lambda}\bm{\mu}}(\bm{\sigma},\bm{\sigma}^\prime)&=\sum_k\log\bigg(1+\text{exp}\bigg[\frac{1}{2} \bm{U}^{[k]}_{\bm{\lambda}}(\bm{\sigma}+\bm{\sigma}^\prime)\\
&+\frac{i}{2}\bm{U}^{[k]}_{\bm{\mu}}(\bm{\sigma}-\bm{\sigma}^\prime)+\bm{d}_{\bm{\lambda}}^{[k]}\bigg]\bigg).
\end{split}
\end{equation}
Note in particular, that the two weight matrices $\bm{U}_{\bm{\lambda}}$ and $\bm{U}_{\bm{\mu}}$ encode the mixing of the physical system with the auxiliary system. In the case where both are set to zero, the state $\psi_{\bm{\theta}}(\bm{\sigma},\bm{a})$ becomes separable and the resulting NDO describes a pure state. 

Before we turn to the machine learning procedure that allows us to reconstruct a physical state, let us further examine the RBM parametrization of the density matrix.
First, note that given a NDO $\bm{\rho}_{\bm{\lambda\mu}}$, it is possible to compute the expectation value of any observable $\bm{\mathcal{O}}$ acting on the physical degrees of freedom $|\bm{\sigma}\rangle$, provided its matrix representation $\mathcal{O}_{\bm{\sigma}\bm{\sigma}^\prime}$ is sparse in that basis (i.e. the number of nonzero elements scales subexponentially with $N$). This can be done simply by considering the observable $\bm{\mathcal{O}}\otimes \bm{I}_{\bm{a}}$ on the composite system:
\begin{equation}
\begin{split}
\langle\bm{\mathcal{O}}\rangle&=\text{Tr}_{\bm{\sigma}}\{\bm{\rho}_{\bm{\lambda\mu}}\bm{\mathcal{O}}\}=\langle\psi_{\bm{\lambda\mu}}|\bm{\mathcal{O}}\otimes \bm{I}_{\bm{a}}|\psi_{\bm{\lambda\mu}}\rangle\\
&=\sum_{\bm{\sigma}\bm{\sigma}^\prime}\sum_{\bm{a}}\psi_{\bm{\lambda\mu}}(\bm{\sigma},\bm{a})\psi^*_{\bm{\lambda\mu}}(\bm{\sigma}^\prime,\bm{a})\mathcal{O}_{\bm{\sigma}^\prime\bm{\sigma}}\\
&=\sum_{\bm{\sigma}\bm{a}}|\psi_{\bm{\lambda\mu}}(\bm{\sigma},\bm{a})|^2\sum_{\bm{\sigma}^\prime}\frac{\psi^*_{\bm{\lambda\mu}}(\bm{\sigma}^\prime,\bm{a})}{\psi^*_{\bm{\lambda\mu}}(\bm{\sigma},\bm{a})}\mathcal{O}_{\bm{\sigma}^\prime\bm{\sigma}}
\end{split}
\end{equation}
Therefore, one can approximate the expectation value of $\bm{\mathcal{O}}$ with a Monte Carlo average of the observable 
\begin{equation}
\mathcal{O}_L(\bm{\sigma},\bm{a})=\sum_{\bm{\sigma}^\prime}\sqrt{\frac{p_{\bm{\lambda}}(\bm{\sigma}^\prime,\bm{a})}{p_{\bm{\lambda}}(\bm{\sigma},\bm{a})}}e^{i(\phi_{\bm{\mu}}(\bm{\sigma},\bm{a})-\phi_{\bm{\mu}}(\bm{\sigma}^\prime,\bm{a}))}\mathcal{O}_{\bm{\sigma}^\prime\bm{\sigma}}
\label{local_O}
\end{equation}
over a collection of samples drawn from the distribution $|\psi_{\bm{\lambda\mu}}(\bm{\sigma},\bm{a})|^2=Z_{\bm{\lambda}}^{-1}p_{\bm{\lambda}}(\bm{\sigma},\bm{a})$. The sparsity of $\mathcal{O}_{\bm{\sigma}\bm{\sigma}^\prime}$ ensures that we can perform the summation in Eq.~(\ref{local_O}) efficiently.
This type of sampling is natural in an RBM (a stochastic neural network) because of its special architecture with edges connecting units between different layers only.
One can show that sampling the distribution $p_{\bm{\lambda}}(\bm{\sigma},\bm{a})$ is equivalent to sampling the conditional distributions $p_{\bm{\lambda}}(\bm{\sigma}\:|\:\bm{h},\bm{a})$, $p_{\bm{\lambda}}(\bm{h}\:|\:\bm{\sigma})$ and $p_{\bm{\lambda}}(\bm{a}\:|\:\bm{\sigma})$, which do not require the knowledge of the normalization constant. Furthermore, each of these conditional distributions factorizes over the unit of the corresponding layer, thus enabling one to sample all the units simultaneously. 

{\it Quantum state reconstruction.} Let us now consider the problem of reconstructing an unknown quantum state $\bm{\varrho}$ from a set of experimental measurements.
In contrast to other quantum state tomography techniques, which extract the elements of the density matrix from the averages of a set of measured observables, we consider instead a collection of raw density measurements $\bm{\sigma}^{\bm{b}}=(\sigma_1^{b_1},\dots,\sigma_N^{b_N})$ in a set of $N_b$ bases $\bm{b}=(b_1,\dots,b_N)$. Given a basis $\bm{b}$, the measurements are distributed according to the probability distribution $P(\bm{\sigma}^{\bm{b}})=\varrho(\bm{\sigma}^{\bm{b}},\bm{\sigma}^{\bm{b}})$. The goal for the training of the neural network is then to find the set of parameters $(\bm{\lambda}^*,\bm{\mu}^*)$ such that the NDO approximates the target density matrix $\bm{\rho}_{\bm{\lambda}^*\bm{\mu}^*}\sim\bm{\varrho}$. The optimal values are discovered by minimizing the divergence between the probability distributions imposed by $\bm{\rho}_{\bm{\lambda}\bm{\mu}}$ and $\bm{\varrho}$, which is expressed in terms of the sum of Kullbach-Leibler (KL) divergences in each basis,  $\Xi_{\bm{\lambda},\bm{\mu}}=\sum_{\bm{b}}\text{KL}_{\bm{\lambda},\bm{\mu}}({\bm{b}})$, where 
\begin{equation}
\text{KL}_{\bm{\lambda},\bm{\mu}}({\bm{b}})=\sum_{\bm{\sigma}^{\bm{\bm{b}}}}P(\bm{\sigma}^{\bm{b}})\log\frac{P(\bm{\sigma}^{\bm{b}})}{\rho_{\bm{\lambda},\bm{\mu}}(\bm{\sigma}^{\bm{b}},\bm{\sigma}^{\bm{b}})} .
\end{equation}
Rather than performing the average over the distribution $P(\bm{\sigma}^{\bm{b}})$ which is unknown, we approximate $\Xi_{\bm{\lambda},\bm{\mu}}$ by averaging over the experimental available data. Assuming we have data sets $\bm{\mathcal{D}}_{\bm{b}}$ containing $||\bm{\mathcal{D}}_{\bm{b}}||$ snapshots $\bm{\sigma}^{\bm{b}}$ in various bases $\bm{b}$, the total divergence becomes $\Xi_{\bm{\lambda},\bm{\mu}}\sim\mathcal{H}(P)+\langle\mathcal{L}_{\bm{\lambda},\bm{\mu}}\rangle$, where $\mathcal{H}(P)\propto\sum_{\bm{b}}\langle P\log P\rangle_{\mathcal{D}_{\bm{b}}}$ is a constant entropy term, and 
\begin{equation}
\langle\mathcal{L}_{\bm{\lambda},\bm{\mu}}\rangle=-\sum_{\bm{b}}||\bm{\mathcal{D}}_{\bm{b}}||^{-1}\sum_{\bm{\sigma}^{\bm{\bm{b}}}\in\bm{\mathcal{D}}_{\bm{b}}}\log \rho_{\bm{\lambda},\bm{\mu}}(\bm{\sigma}^{\bm{b}}_k,\bm{\sigma}^{\bm{b}}_k)
\label{NLL}
\end{equation}
is the negative log-likelihood averaged over the data, relevant for the optimization.
Each iteration of the training consists of updating the network parameters $\bm{\theta}$ according to an optimization algorithm, the simplest one being stochastic gradient descent:
\begin{equation}
\bm{\theta}\leftarrow\bm{\theta}-\eta\nabla_{\bm{\theta}}\langle\mathcal{L}_{\bm{\lambda},\bm{\mu}}\rangle_{\bm{\mathcal{D}}_\ell}
\end{equation}
where the gradient step $\eta$ is called learning rate, and the average negative log-likelihood is estimated over a random subset of training samples $\bm{\mathcal{D}}_\ell\in\bigcup_{\bm{b}}\bm{\mathcal{D}}_{\bm{b}}$.

In order to take the derivative of Eq.~\eqref{NLL}, we first need to rotate the density operator back into the original reference basis $\bm{\sigma}$ via the relation $\bm{\rho}_{\bm{\lambda},\bm{\mu}}^{\bm{b}}=\bm{\mathcal{U}}_{\bm{b}}\bm{\rho}_{\bm{\lambda},\bm{\mu}}\bm{\mathcal{U}}_{\bm{b}}^\dagger$. The matrix  $\bm{\mathcal{U}}_{\bm{b}}$ is simply given by the product of unitary matrices $\mathcal{U}_{\bm{b}}(\bm{\sigma}^{[\bm{b}]},\bm{\sigma})=\bigotimes_j \bm{\mathcal{U}}_{b_j}$, each performing a local change of basis $\bm{\mathcal{U}}_{b_j}=\langle\sigma_j^{b_j}|\sigma_j\rangle$ \cite{torlai_Tomo}. The gradients of the average negative log-likelihood $\langle\mathcal{L}_{\bm{\lambda},\bm{\mu}}\rangle_{\bm{\mathcal{D}}}$ with respect to the network parameters become
\begin{equation}
\begin{split}
\nabla_{\bm{\lambda}}\langle\mathcal{L}_{\bm{\lambda},\bm{\mu}}\rangle_{\bm{\mathcal{D}}}&=-\sum_{\bm{b}}||\bm{\mathcal{D}}_{\bm{b}}||^{-1}\sum_{\bm{\sigma}^{\bm{\bm{b}}}\in\bm{\mathcal{D}}_{\bm{b}}}\langle\nabla_{\bm{\lambda}}\Gamma^{[+]}_{\bm{\lambda}}+\nabla_{\bm{\lambda}}\Pi_{\bm{\lambda}\bm{\mu}}\rangle_{\bm{Q}_{\bm{\sigma}^{\bm{b}}}}\\
&+\langle\nabla_{\bm{\lambda}}\log \tilde{\rho}_{\bm{\lambda},\bm{\mu}}(\bm{\sigma},\bm{\sigma})\rangle_{\bm{\rho}_{\bm{\lambda},\bm{\mu}}}\\
\end{split}
\label{der_lambda}
\end{equation}
and
\begin{equation}
\nabla_{\bm{\mu}}\langle\mathcal{L}_{\bm{\lambda},\bm{\mu}}\rangle_{\bm{\mathcal{D}}}=-\:\sum_{\bm{b}}||\bm{\mathcal{D}}_{\bm{b}}||^{-1}\sum_{\bm{\sigma}^{\bm{\bm{b}}}\in\bm{\mathcal{D}}_{\bm{b}}}\langle i \nabla_{\bm{\mu}}\Gamma^{[-]}_{\bm{\mu}}+\nabla_{\bm{\mu}}\Pi_{\bm{\lambda}\bm{\mu}}\rangle_{\bm{Q}_{\bm{\sigma}^{\bm{b}}}}
\label{der_mu}
\end{equation}
The averages 
\begin{equation}
\langle \bm{\mathcal{O}}\rangle_{\bm{Q}_{\bm{\sigma}^{\bm{b}}}}=
\frac{\sum_{\bm{\sigma}\bm{\sigma}^\prime}Q_{\bm{\sigma}^{\bm{b}}}(\bm{\sigma},\bm{\sigma}^\prime)\mathcal{O}(\bm{\sigma},\bm{\sigma}^\prime)}{\sum_{\bm{\sigma}\bm{\sigma}^\prime}Q_{\bm{\sigma}^{\bm{b}}}(\bm{\sigma},\bm{\sigma}^\prime)}.
\end{equation}
with respect to the quasiprobability distributions $Q_{\bm{\sigma}^{\bm{b}}}(\bm{\sigma},\bm{\sigma}^\prime)=\mathcal{U}_{\bm{b}}(\bm{\sigma}^{\bm{b}},\bm{\sigma})\rho_{\bm{\lambda},\bm{\mu}}(\bm{\sigma},\bm{\sigma}^\prime)\mathcal{U}^*_{\bm{b}}(\bm{\sigma}^{\bm{b}},\bm{\sigma}^\prime)$
can be evaluated directly on the samples in the data sets, with the double summation running over $4^t$ terms for a basis $\bm{b}$ where there are only $t$ local unitaries $\bm{\mathcal{U}}_{b_j}\ne\bm{I}_j$. On the other hand, the average of the log-probability over the full-model probability distribution $\langle\nabla_{\bm{\lambda}}\log \tilde{\rho}_{\bm{\lambda},\bm{\mu}}(\bm{\sigma},\bm{\sigma})\rangle_{\bm{\rho}_{\bm{\lambda},\bm{\mu}}}$ appearing in Eq.~(\ref{der_lambda}) requires the knowledge of the normalization constant $Z_{\bm{\lambda}}$ and can be computed exactly only for very small system sizes. 
For larger $N$, it is possible to approximate this average by running a Markov-chain Monte Carlo simulation on the distribution $\rho_{\bm{\lambda},\bm{\mu}}(\bm{\sigma},\bm{\sigma})$. 
Instead of reaching equilibrium at each training iteration, the chain is initialized with a training sample and statistics are collected after few sampling steps, resulting into a fast learning procedure. The algorithm, called contrastive divergence \cite{CDk}, has been widely used for unsupervised pretraining of large, deep neural networks~\cite{Hinton06,Salakhutdinov08,Salakhutdinov09}.

\begin{figure}[t]
\noindent \centering{}\includegraphics[width=1\columnwidth]{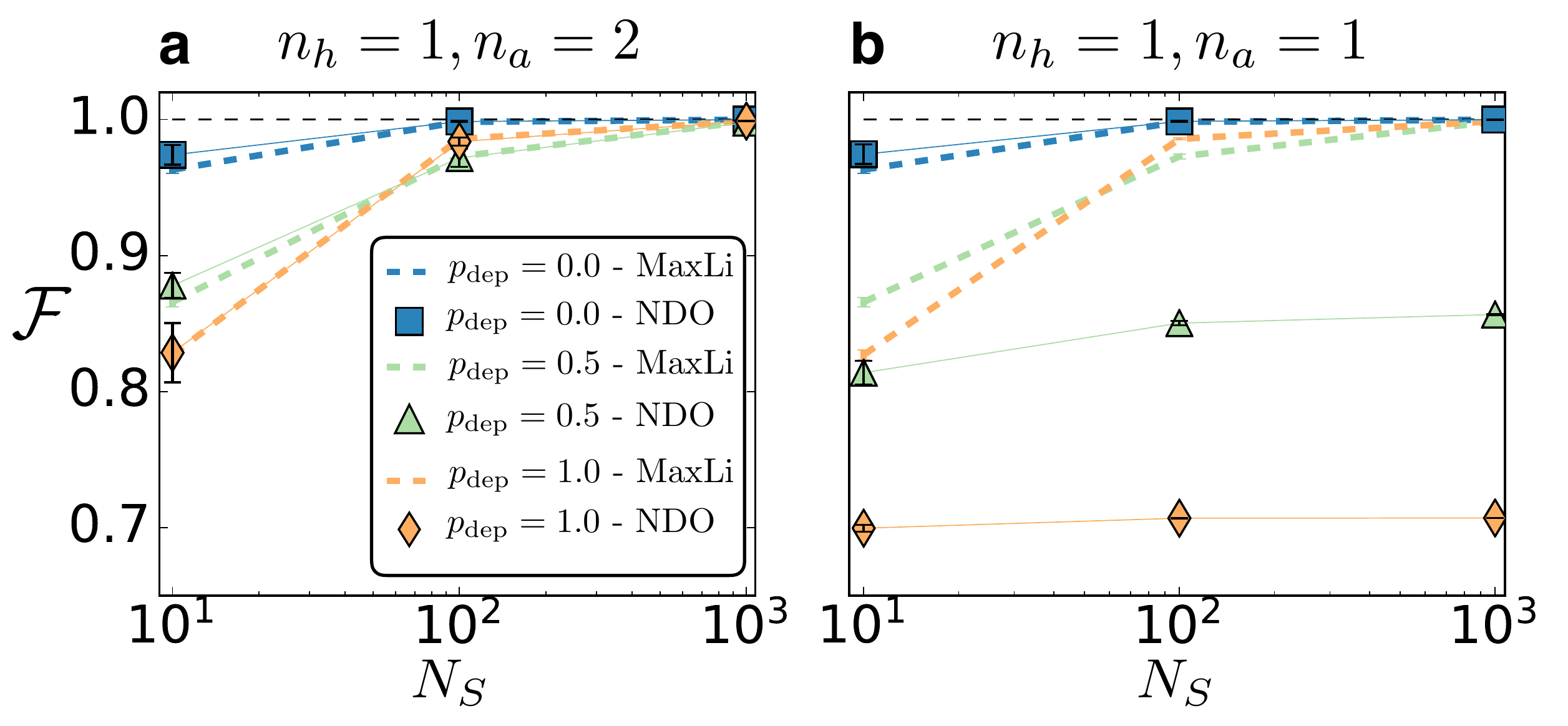}
\caption{Comparison of the reconstruction fidelities between NDO and MaxLi tomography for a Bell state $|\Phi^+\rangle$ undergoing a depolarizing channel with strength $p_{\text{dep}}$. We show the scaling of the fidelity as a function of the number of measurements per basis $N_S$ for two different choices of network structure (each point is plotted with standard deviation error from an average over 100 realizations of the data set).}
\label{Fig2} 
\end{figure}

{\it Results.} Let us now demonstrate the NDO parametrization and reconstruction for entangled photonic states, focusing on small systems where the problem is tractable.
The tomographic reconstruction of the density matrix for such states is widely used in a variety of tasks. These include the characterization of optical processes~\cite{Hamel14}, detectors~\cite{Grandi17}, and the tests of local realism of quantum mechanics~\cite{Lavoie09,Angelo06}. We consider the case of two qubits, setting the number of hidden units to $n_h=1$, and initialize the weights with a uniform distribution centered around zero with width $w=0.01$ (and biases set to zero). The network parameters are updated using the AdaDelta optimization algorithm~\cite{Zeiler2012} over training batches containing 10 samples, and the best network is discovered by choosing $(\bm{\lambda}^*,\bm{\mu}^*)$ for which the average log-likelihood is maximum. We quantify the performance of the reconstruction by computing the fidelity between $\bm{\rho}_{\bm{\lambda}^*\bm{\mu}^*}$ and the target density operator $\bm{\varrho}$, defined as $\mathcal{F}=\text{Tr}\{\sqrt{\sqrt{\bm{\rho}_{\bm{\lambda}^*\bm{\mu}^*}}\bm{\varrho}\sqrt{\bm{\rho}_{\bm{\lambda}^*\bm{\mu}^*}}}\}$. 

We first consider the ideal situation where the only fluctuations in the measurement outcomes are of statistical nature.  Thus, we generate a synthetic data set using the exact target quantum state $\bm{\varrho}$. We choose to reconstruct the Bell state $|\Phi^+\rangle=(|00\rangle+|11\rangle)/\sqrt{2}$ undergoing a depolarizing channel, where we introduce a controllable amount of mixing through the channel strength $p_{\text{dep}}$. The mixed state is described by the density matrix $\bm{\varrho}=(1-p_{\text{dep}})|\Phi^+\rangle\langle\Phi^+| + p_{\text{dep}}\bm{I}/4$,
where we have the pure state $|\Phi^+\rangle$ for $p_{\text{dep}}=0$ and the maximally mixed state $\bm{I}/4$ for $p_{\text{dep}}=1$. We build the data sets by measuring the system in the $N_b=9$ bases $\{\sigma_0^i,\sigma_1^j\}$ with $j=x,y,z$. Further, we generate multiple data sets with a different number $N_S$ of measurements per basis (each containing 2 bits of information). We report in Fig.~\ref{Fig2} the fidelities obtained after training the NDO for three different depolarizing strengths and an increasing number of measurements per basis.
We compare our results with the fidelities obtained with standard maximum likelihood (MaxLi) tomography~\cite{Banaszek99,James01}. We observe slightly better fidelities when using two auxiliary units (Fig.~\ref{Fig2}a), while the NDO with $n_a=1$ is not capable of purifying the state of the physical system (Fig.~\ref{Fig2}b).

\begin{figure}[t]
\noindent \centering{}\includegraphics[width=0.9\columnwidth]{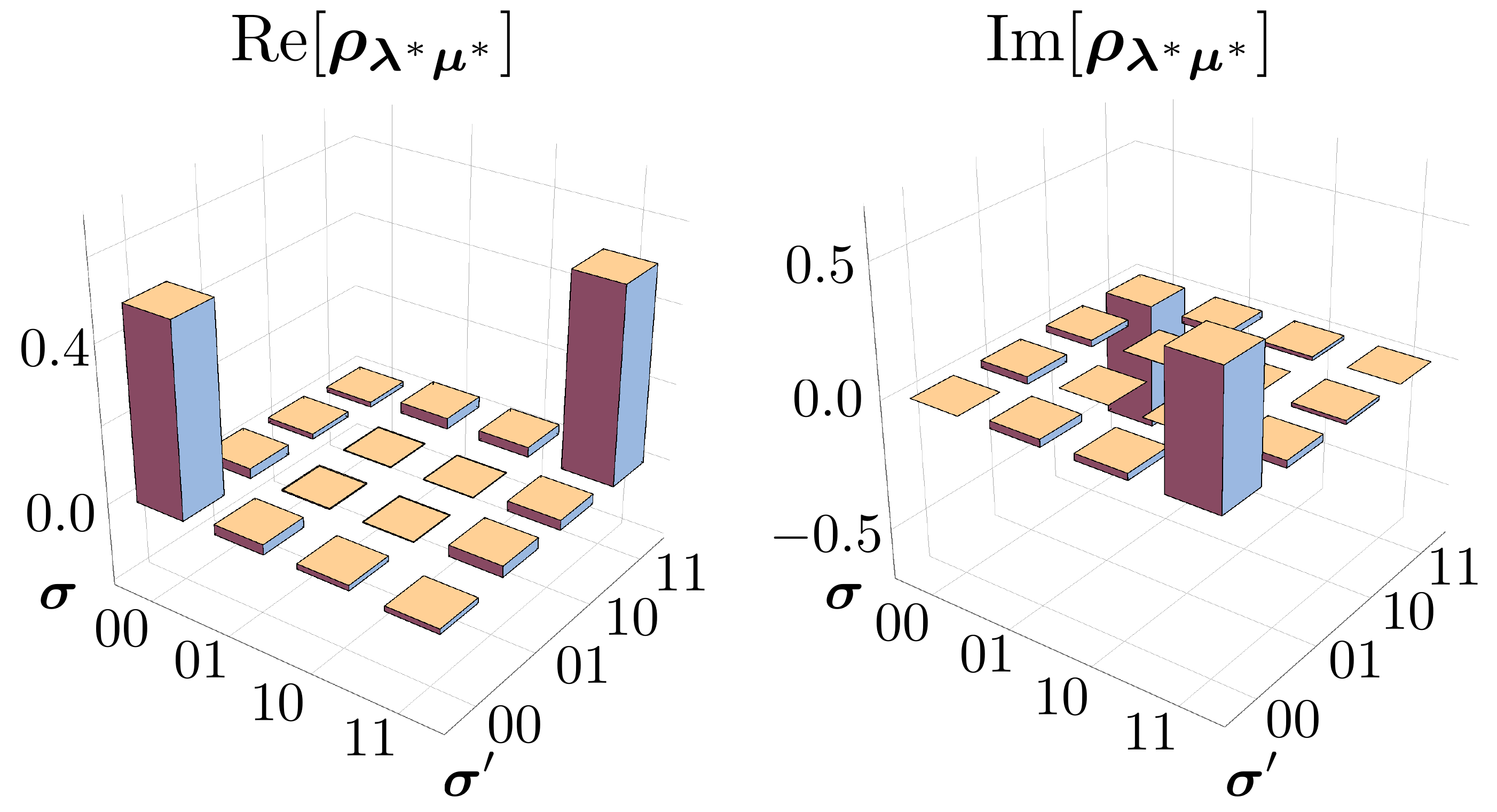}
\caption{Real and imaginary parts of the reconstructed NDO, trained on experimental coincidence counts for the two-qubit state $|\Psi\rangle=\frac{1}{\sqrt{2}}(|00\rangle+i|11\rangle)$.}
\label{Fig3} 
\end{figure}

Finally, we consider real experimental data, where unknown sources of noise are present. Using the coincidence counts provided by Ref.~\cite{Alteper2006}, we perform NDO tomography on the experimental measurements for the state $|\Psi\rangle=\frac{1}{\sqrt{2}}(|00\rangle+i|11\rangle)$, where the degrees of freedom represent the polarizations of the entangled photons. In Fig.~\ref{Fig3} we plot the real and imaginary parts of the reconstructed NDO, selected with the same criterion of minimum negative log-likelihood.  The fidelity between the NDO and the ideal state is found to be $\mathcal{F}_{\text{NDO}}=0.9976$, with MaxLi tomography achieving similar fidelity $\mathcal{F}_{\text{MaxLi}}=0.992$.

{\it Conclusions.}  
We have devised and constructed a machine learning algorithm, based on a restricted Boltzmann machine, that is capable 
of storing a representation, and performing generative modeling, on a quantum state with arbitrary mixing.  The resulting graphical model
purifies the mixed state by enlarging the Hilbert space with the use of latent, or hidden, units in the stochastic neural network.
The model can be readily trained by standard machine learning techniques, including contrastive divergence, with measurements from an
arbitrary basis, thereby allowing approximate quantum tomography to be performed on any mixed state.  We demonstrate the technique on typical
two-photon entangled states, including real experimental data with unknown noise sources, and achieve fidelities competitive with standard tomographic
techniques.

As machine learning techniques continue to become integrated into the field of quantum information science and technology,
we anticipate their role in error correction, state and process tomography, and other tasks in validation will rapidly increase.
Restricted Boltzmann machines offer a powerful method for generative modeling, with training algorithms that are well studied by the 
machine-learning community. Their demonstrated ability to provide practical tradeoffs between representation, computation, and statistics, offers a rich field of study in the case of quantum states, which will be important in the integration of classical and quantum algorithms inevitable in near-term devices and computers.

\subsection*{Acknowledgements }

We thank L. Aolita, M. Beach, G. Carleo, J. Carrasquilla, M. Endres, B. Kulchytskyy, J.-P.~MacLean, K. Resch, S. Weinstein and E. van Nieuwenburg for useful discussions. This research was supported by NSERC, the CRC program, the Ontario Trillium Foundation, the Perimeter Institute for Theoretical Physics, and the National Science Foundation under Grant No. NSF PHY-1125915. Simulations were performed on resources provided by SHARCNET. Research at Perimeter Institute is supported through Industry Canada and by the Province of Ontario through the Ministry of Research $\&$ Innovation.

\bibliographystyle{apsrev4-1}
\bibliography{Bib.bib}

\end{document}